\documentclass[aps,prl,twocolumn,superscriptaddress,amssymb]{revtex4}
\usepackage{graphicx}

\def\simle{\mathrel{
   \rlap{\raise 0.511ex \hbox{$<$}}{\lower 0.511ex \hbox{$\sim$}}}}

\def\simge{\mathrel{%
    \rlap{\raise 0.511ex \hbox{$>$}}{\lower 0.511ex \hbox{$\sim$}}}}

\def\be{\begin{eqnarray}}
\def\ee{\end{eqnarray}}

\begin{document}

\title{Large $N_c$ confinement and turbulence}

\author{Jean-Paul Blaizot}
\email{Jean-Paul.Blaizot@cea.fr}
\affiliation{ECT$^*$, Strada delle Tabarelle 286, I-38050 Villazzano
  (Trento), Italy}

\author{Maciej A. Nowak}
\email{nowak@th.if.uj.edu.pl}
\affiliation{M. Smoluchowski Institute of Physics and
Mark Kac Center for Complex Systems Research,
Jagiellonian University, PL--30--059 Cracow, Poland\\
GSI, Planckstr. 1, 64291 Darmstadt, Germany}

\date{\today}

\setlength{\parindent}{2ex}
\setlength{\parskip}{1ex plus 0.5ex minus 0.2ex}


\begin{abstract}
We suggest that the transition  that occurs at large $N_c$ in the
eigenvalue distribution of a Wilson loop may have a turbulent origin. We
arrived at this conclusion by  studying the complex-valued inviscid
Burgers-Hopf equation that corresponds to the Makeenko-Migdal loop
equation, and we demonstrate the appearance of a shock  in the
spectral flow of the Wilson loop eigenvalues. This picture
supplements that of the
 Durhuus-Olesen  transition with a particular realization
of disorder. The critical behavior at the formation of the shock
allows us to infer exponents  that have been measured recently in
lattice simulations by Narayanan and Neuberger in $d=2$  and $d=3$.
Our analysis leads us to speculate that the universal behavior
observed in these lattice simulations might be a generic feature of
confinement, also in $d=4$ Yang-Mills theory.

\end{abstract}



\maketitle


Many efforts continue to be devoted to the study of QCD in the limit
of a large number of colors, after the initial suggestion by t'Hooft
\cite{THOOFT}. This, in part, is due to the general belief that the
large $N_c$ limit captures the essence of confinement, one of the
most elusive of QCD properties. At the same time  the theory
simplifies considerably in the large $N_c$ limit: fluctuations die
out and  the measure of the partition function becomes localized at
one particular configuration, making the large $N_c$ limit akin to a
classical approximation  \cite{MASTER}.

This paper will discuss the large $N_c$ limit of Yang-Mills theory
in 2 dimensions, but we have good reasons to believe that much of
our analysis can be extended to higher dimensions. There are many
equivalent approaches to multicolor Yang-Mills in $d=2$. For
definiteness, we shall refer to the
 known~\cite{GOPAKUMAR}
 formulation in terms of
free random variables~\cite{VOICULESCU,SPEICHER}. This translates
$d=2$ Yang-Mills theory onto large $N_c$ matrix model, where the
size of the unitary matrix is identified with the number of colors.
More specifically, the basic observable that we shall consider is
the Wilson loop along a (simple) curve $C$  \be W(A)=\left<P\exp
i\oint_C A_{\mu}dx_{\mu}\right>, \label{Wloop} \ee where the
averaging is over the Yang-Mills  measure, and we have made explicit
that $W$ depends in fact only on the area $A$ enclosed by $C$, to
within a normalization \cite{DURHUUS}. The matrix $W$ is unitary,
with eigenvalues of the form $\lambda= \exp(i\theta)$ that are
distributed, in the limit $N_c\to\infty$, according to an average
density $\rho(\theta,A)$.

Recently,  Narayanan and Neuberger~\cite{NARNEU}
 studied the behavior
of $W(A)$ as a function of $A$ in the large $N_c$ limit. They
observed that for small loops (which probe short distance,
perturbative phyics),  the spectrum does not cover the whole unit
circle, but exhibits a gap; in contrast,   for very  large loops
(which probe long distance, nonperturbative physics)  the spectrum
covers uniformly the unit circle (gapless phase). Since the
crossover region is becoming infinitely thin~\cite{NARNEU} in  the
limit $N_c \rightarrow \infty$, one is tempted to try and explain
the transition using  classical concepts only, which is precisely
what we aim at in this note. In fact this behavior of the spectrum
agrees with the order(gapped)-disorder(gapless)  transition,
proposed long ago in the context of large $N_c$ two-dimensional Yang-Mills
theory by Durhuus and Olesen~\cite{DURHUUS} and based on the
explicit solutions
 of corresponding Makeenko-Migdal equations~\cite{MM}. Surprisingly, a
 similar critical behavior has been  observed   also in
 $d=3$ dimensions and conjectured to hold in $d=4$ large $N_c$
 Yang-Mills theory~\cite{NARNEU}.
In this letter we suggest the general mechanism for such a
transition, by tracing the complex singularities of the eigenvalue
flow equation,  and demonstrating the appearance of  ``spectral
shocks'' that may signal the transition to a turbulent state.

The spectral density  is  not available in  analytic form, but the moments \be
w_n(A)\equiv <{\rm tr} W(A)^n> =\int_{-\pi}^{+\pi} d\theta \,{\rm
e}^{i n\theta} \rho(\theta, A)\label{moments}\ee are.
 An explicit, compact form for these moments is given in Ref.~\cite{GOPAKUMAR}  in terms of an integral
representation \be \label{moments2}w_n(A)&=&\frac{1}{n} \oint \frac{dz}{2\pi i}
(1+1/z)^n \exp(-nA(z+1/2))
\nonumber \\
&=&\frac{1}{n} L_{(n-1)}^1(nA) \exp (-nA/2) \ee where the
representation of Laguerre polynomials, used in the second line,
allows  connection to results  known already 25 years
ago~\cite{DURHUUS,VAST}. The Durhuus-Olesen transition can be seen by studying the
asymptotic behavior of these   Laguerre polynomials, using
 a saddle point analysis of their integral representation ~\cite{MATYTSIN,OLESEN}. The result is
surprising:
 for  a loop area
below the critical value $A_c=4$, the moments oscillate and decay
like $n^{-3/2}$, while for  $A>A_c$  the moments decay exponentially
with $n$, modulo similar power behavior.
 Both regimes are separated by double scaling
limit. There exists a vast literature on this subject~\cite{VAST}.
Here we would only like to stress, that the transition is subtle,
and popular arguments that two-dimensional confinement is trivial,
and perturbative (linear potential),  can easily lead to paradoxes
when, e.g., instanton effects are not taken into
account~\cite{VAST}.

In order to analyze the nature of the transition, we   consider the following function
 \cite{GOPAKUMAR} \be F( \theta,A)= i\left({\rm
e}^{i\theta}G({\rm e}^{i\theta},A)-\frac{1}{2}\right),  \ee where $G(z)$ is
the resolvent \be G(z,A)=\int_{-\pi}^{+\pi} d\alpha
\,\frac{\rho(\alpha,A)}{z-{\rm e}^{i\alpha}}. \ee  The
function $F(\theta,A)$ is analytic in the  complex $\theta$-plane, with a
discontinuity  across the real axis proportional to the spectral
density, $\rho(\theta,A)=(1/\pi){\rm Im} F(\theta-i0^+,A)$. A simple
calculation yields the explicit expressions \cite{DURHUUS}
\be
\label{resolvent2} F(\theta,A)&=&i\left( \frac{1}{2}+
\sum_{n=1}^{+\infty}   w_n(A) {\rm e}^{-in\theta} \right)\nonumber\\
&=& \frac{1}{2}\int_{-\pi}^{+\pi} d\alpha\,
\rho(\alpha,A)\cot\left(\frac{\theta-\alpha}{2}\right) ,
\label{Resolvent}\ee whose
imaginary part gives the spectral density in the form
 \be \rho(\theta,
A)= \frac{1}{2\pi}\left( 1+\sum_{n=1}^{+\infty}2w_n(A)
\cos(n\theta)\right) ,\label{spectral}\ee in agreement with
Eq.~(\ref{moments}) above (we have used the fact that the moments
are real and that $\rho(-\theta)=\rho(\theta)$).

It can be shown that the function $F$ obeys the following equation
\cite{GOPAKUMAR,DURHUUS,VAST} \be
\partial_AF+F\partial_{\theta}F=0
\label{ComplexBurgers} .\ee This is the so-called complex Burgers
equation in the inviscid limit. This equation is analogous to the
real Burgers equation of fluid dynamics (with $A$ playing the role
of time, $\theta$ that of  a coordinate, and $F$ of a velocity field).
The complex Hopf-Burgers   equation  is omnipresent in Free Random
Variables calculus~\cite{VOICULESCU}. This, and similar integro-differential equations, also appear
frequently as   one dimensional models for quasi-geostrophic
equations, describing e.g. the dynamics of the mixture of cold and
hot air and the fronts between them. Another reason why the complex
Burgers equation is much studied  is the fact that the structure of
quasi-geostrophic equations resembles 3-d Euler equations, if one
substitutes the velocity by the vorticity~\cite{MAJDA2}.  Here, we
shall take advantage of the abundant mathematical studies of the
complex Burgers equation to suggest a connection between signals
of turbulence (commonly associated with the blow-up of the solution in finite time)
and the problem of spectral flow of the eigenvalues of
Wilson loop operators. We shall in particular   adapt the proof
 of the blow-up from Ref.~\cite{MORLET}, using the method of
complex characteristics, and tracing singularities in the complex plane.
 The observation of  shock formation in the complex Burgers and similar
equations was confirmed by other methods, see e.g.~\cite{CHAE}.

In the present case, the  method of characteristics provides the
following implicit solution in terms of an auxiliary  function
$\xi(A,\theta)$: \be F(A,\theta)= F_0(\xi(A,\theta)),\qquad
\theta=\xi+ A F_0(\xi),  \ee where  $F_0$ is given by the initial
condition  \be \label{initialF0} F_0(\theta)=F(A=0,\theta) . \ee
This implicit solution is well defined as long as the mapping
between $\theta$ and $\xi$ remains regular. However singular points
occur when $d\theta/d\xi=0$, that is, for $\xi_c$ solution  of
\be\label{criticalcond} 1+AF_0'(\xi_c)=0.\ee A lot of information on
the solution, and in particular the occurrence of a blow up, can be
inferred from the study of the location of these singular points.

We begin our analysis with  the  ordered (gapped) state, and recover
 known results (see e.g. the second Ref.~\cite{OLESEN}) in an easy and  direct way.
 We start from  an initial
 condition peaked at eigenvalue $\lambda =1$, i.e.
$\rho(\theta,0)=\delta (\theta)$ (corresponding to $A=0$ in Eq.~(\ref{moments2})). Then, from
(\ref{Resolvent}),  $F_0(\theta)=1/2 \cot \theta/2$,
and from (\ref{criticalcond}) $ \sin^2\xi_c/2=A/4$.
In the vicinity of the singularity the characteristics behave as
\be\label{localanal}
\theta=\theta_c+\frac{1}{4}(\xi-\xi_c)^2\sqrt{1/A-4}
\ee
where  $
\theta_c=\xi_c+\sqrt{A(1-A/4)}$
and  $\xi_c= 2 \arcsin (\sqrt{A}/2)$. The spectral density is
easily deduced from the imaginary part of $F$ or equivalently $F_0$ (see (\ref{initialF0})).
A simple analysis then reveals
 that $\rho(\theta,A)=0$ when $\theta>\theta_c$,
  while for $\theta\simle \theta_c$, $\rho(\theta,A)\sim \sqrt{\theta_c-\theta}$.
  In other words, $\theta_c$ determines the edge of the spectrum which, as long as $A<4$ presents a gap.

At  the closure of the gap, i.e., for $A=4$, the second derivative in the
expansion (\ref{localanal}) vanishes,  and we have instead
\be\label{cubic}
\theta=\pi+\frac{1}{A}(\xi-\xi_c)^3,\ee
with the spectral density $\rho(\theta,A)\sim (\pi-\theta)^3$ for $\theta\simle\pi$.

From this point on,  as $A$ keeps increasing, $ \Re \xi_c$ remains equal
to $\pi$, but the initial singularity splits into two complex conjugate
ones that move away to $\pm i\infty$, leading eventually to a uniform spectral density at large $A$.
This behavior  is  reminiscent of the turbulent
inverse spectral cascade (alike in two-dimensional turbulence \cite{Kraichnan}) which,
as $A$ grows, {\em suppresses} higher Fourier modes, leaving in the
$A=\infty$ limit
only the longest wavelength mode
corresponding here to the constant density $\rho(\theta)=1/2\pi$.

To demonstrate the inverse spectral cascade, we consider, following
\cite{MORLET}, a small perturbation of the form \be
\rho(\theta,A_0)=\frac{1}{2\pi}(1+2\epsilon \cos \theta)+{\cal O}(\epsilon^2), \ee where
$A_0>>1$. This is of the form (\ref{spectral}) with $w_1=\epsilon$,
and all other moments vanishing exponentially. It follows from (\ref{Resolvent})
that $ F_0(\xi) =\frac{i}{2}(1+2\epsilon \exp (-i\xi))$, $\theta =
\xi +(A-A_0) F_0(\xi)$, and a singularity occurs  when $\exp(i\xi_c)
= -\epsilon (A-A_0)$. In the vicinity of the singularity \be
\theta=\theta_c+\frac{i}{2}(\xi-\xi_c)^2, \ee where
$\theta_c=\xi_c+(A-A_0)F_0(\xi_c)$. We have  two solutions depending
on whether $A>A_0$ or $A<A_0$. In the first case \be
\theta_c=\pi-i\left(1-\frac{A-A_0}{2} +\ln \epsilon(A-A_0) \right),
\ee whereas in the second case \be
\theta_c=-i\left(1+\frac{A_0-A}{2} +\ln \epsilon (A_0-A) \right).
\ee In the first case, the singularity is initially (when $A=A_0$)
at $ \theta_c=\pi +i\infty$, and for small $\epsilon$ ($\epsilon <
1/2$), it remains complex, and returns to $+i\infty$ as $A\to
\infty$: asymptotically the effect of the perturbation vanishes, and
only the constant mode survives.  However in the second case,   the
initial singularity   at $+i\infty$ moves towards the real axis as
$A$ decreases, and reaches it in  a finite time $A^*$  given by \be
0=1+\frac{A_0-A^*}{2} +\ln \epsilon(A_0-A^*). \ee At this point we
have a  blow-up~\cite{MORLET}, of the type already encountered
earlier at the closure of the gap. In fact,  this phenomenon is
quite generically associated to the motion (as $A$ decreases) of
complex conjugate singularities (square root branch points)  towards
the real axis, that eventually eventually merge on the real axis
(into a third order branch point).

This generic phenomenon  has  an exact analog in optics
~\cite{BERRY}, where  light  rays play the role of the   
  characteristics. The singularities lines are there the
caustics, and the merging of singularities discussed above
corresponds to the  merging of two ``fold'' caustics into a ``cusp''
(in the terminology of catastrophe theory). The essence of the cusp
singularity is captured by the  Pearcey's function \be P(\xi,
\eta)=\int_{-\infty}^{\infty} dt \exp (i(f^4/4+\xi t^2/2 +\eta t))
\ee whereas the square root type singularity is described by the well
known Airy function. This analogy can be used in order to explain
the origin of the universal function for Yang-Lee zeroes for $d \ge
2$ dimensional Yang-Mills obtained in Ref.~\cite{NARNEU}. The explicit
comparison will be presented elsewhere~\cite{BLAIZOT2}.

The location of the singularities in the complex plane also
determine the asymptotic behavior of the moments, according to
general arguments ~\cite{FRISCH}.  When $\theta_c$ is close  to the
real axis, one may write $\theta_c \equiv \theta^* +i\Delta(A) $,
with $\theta^*$ real; then the singularity is of the form
$(\theta-\theta^*)^{\mu}$ , where $\mu$ is not an integer (branch
point singularity).  It follows that the large $n$ Fourier
coefficients, i.e., the moments $w_n$,  behave as \be w_n=
|n|^{-(\mu+1)}e^{-n\Delta(A)} {\Re} e^{i n \theta^*} .\ee Thus, the
position of the  singularity determines the width of an analytical
strip, controlled by the value of $\Delta(A)$.  In the gapped phase,
the singularity is of the square root type, but always on the real
axis. Then, since ${\Delta=0}$, the moments only oscillate,
accompanied by power law $n^{-3/2}$. In the gapless phase, the same
power law is   accompanied by exponential damping, since in this
case the singularity is
 complex (with a non vanishing $\Delta(A)$).
At the critical point, a generic cubic singularity (see Eq.~(\ref{cubic}) and~\cite{JANWIECZ,JANIKNOWAK})
appears in a  narrow interval where
 $1<< n << 1/\Delta(A)$  (just approaching the singularity from above)
and  the scaling goes like
\be
w_n \sim |n|^{-4/3}.
\ee

This generic picture agrees with detailed  calculations using the
analytic form of the moments given above, based on asymptotics of
Laguerre polynomials~\cite{MATYTSIN,NARNEU,OLESEN}. We would like to
stress however that the method of tracing complex singularities is
general and therefore we expect it to hold also in higher
dimensions. The ``speed" with which the singularities move may
depend on the dimensionality, and possibly on other features of the
loops, but for large loops it is natural to expect that the dominant
control parameter will remain the area of the loop. We may therefore
speculate that,  also in higher dimensions, the disordered phase may be  caused by an
inverse spectral cascade.

One may also argue that the universal nature of the critical
behavior, conjectured and observed in Ref.~\cite{NARNEU}, may find
its origin in the fact, that since the critical behavior happens in
a very narrow  analyticity strip,  one can  expect universal (ergodic) behavior, alike in several models of disorder in mesoscopic
physics. If this is the case, one may expect that  simple schematic
models (e.g. matrix models) may define this  class of universality.
This seems to be indeed the case. A particular illustration is
provided by the   matrix model proposed by Janik and
Wieczorek~\cite{JANWIECZ}, hereafter JW model. The model stems from
the general construction of multiplicative free
evolution~\cite{NOWAK}, where  increments are mutually free in the
sense of Voiculescu. The unitary realization in JW model corresponds
to matrix value unitary random walk, where the evolution operator is
the ordered string of consecutive multiplications of infinitely
large unitary matrices \be W=\left< \prod_i^K U_i \right> \ee where
  $U_i =\exp i\sqrt{t/K}H_i$, with $H_i$ a hermitian random
matrix.  The model is   a random matrix generalization of the
multiplicative random walk performed in $K$ steps during ``time"
$t$. In the continuum limit $K \rightarrow \infty$, the model is
exactly solvable. The solution coincides exactly with the
two-dimensional QCD, provided   one identifies $t$ with the area of
the Wilson loop, modulo a normalization ~\cite{COMMENT}.

Recently, Neuberger and Narayanan~\cite{NARNEU} have observed, that
large $N_c$  Yang-Mills lattice simulations in $d=2$ and $d=3$
demonstrate the same critical scaling at the closure of the gap as
in the JW model and have conjectured that this model establishes a
universality class for $d=4$ large $N_c$ Yang-Mills as well.
 That the whole dynamics
of complicated non-perturbative QCD can be reduced {\it in some
  spectral regime} to matrix model is not new - a notable case is the
universal scaling of  spectral density of Euclidean Dirac operator
for sufficiently small eigenvalues, where the spectrum belongs to
broad universality class of corresponding chiral models~\cite{JAC}.
In fact the present analysis leads us to expect that  in the very
narrow spectral window around $\lambda=-1$ a universal oscillatory
regime  precedes  the formation of the spectral shock, in
qualitative analogy to similar spectral oscillations of the quark
condensate before the spontaneous breakdown of chiral symmetry,
based on  the Banks-Casher relation~\cite{BC}.

In this note, we have proposed to view confinement-deconfinement
transition in multicolor Yang-Mills as a order-disorder phenomenon,
where the transition to disorder  is caused by  (inverse)
turbulence in the spectral flow of Wilson loop operators. This
picture corroborates the picture of order-disorder transition that
Durhuus and Olesen have
 envisioned many years ago,
and supplements it with a detailed model for building the disorder,
based on the development of the inverse spectral cascade. It will be interesting to
incorporate into this picture the effects of finite $N_c$ and matter
fields. We expect, that finite $N_c$ effects will contribute to the
appearance of an effective spectral viscosity $\nu_s$, which will
smoothen the shocks, but will not destroy them~\cite{BLAIZOT2}.


\section*{Acknowledgements}
MAN is grateful to GSI Darmstadt for hospitality during his sabbatical.
This work was  supported by
  Marie Curie TOK Grant  MTKD-CT-2004-517186
"Correlations in Complex Systems" (COCOS).



\end{document}